
\input phyzzx
\nopagenumbers
\voffset = -0.4in
\footline={\ifnum\pageno=1 \nulline \else\newfootline \fi}
\def\nulline{{\hfill}}
\def\newfootline{\advance\pageno by -1\hss\tenrm\folio\hss}
\rightline {October 1993} \rightline {SUSX--TH--93/17}
\rightline {QMW--TH--93/31}
\title {Modular Symmetries in $Z_N$ Orbifold Compactified \break
String Theories
with Wilson Lines
.} \author{D. Bailin$^{a}$, \ A. Love$^{b}$,  \ W.A.
Sabra$^{b}$\ and \ S. Thomas$^{c}$}
\address {$^{a}$School of Mathematical and Physical
Sciences,\break
University of Sussex, \break Brighton U.K.}
\address {$^{b}$Department of Physics,\break
Royal Holloway and Bedford New College,\break
University of London,\break
Egham, Surrey, U.K.}
\address {$^{c}$
Department of Physics,\break
Queen Mary College,\break
University of London,\break
Mile End Road, London,  U.K.}
\abstract { Target space modular symmetries relevant to string loop threshold
corrections
are studied for $Z_N$ orbifold compactified string theories
containing Wilson line background fields.}
\endpage
 \REF\one{L. Dixon, J. A. Harvey, C. Vafa and E. Witten, Nucl. Phys.
B261 (1985) 678; B274 (1986) 285.}
\REF\two{ A. Font, L. E. Ibanez,
F. Quevedo and A. Sierra, Nucl. Phys. B331 (1991) 421.}
\REF\three{R. Dijkgraaf, E. Verlinde and H. Verlinde, Comm. Math.  Phys.115
	      (1988) 649.}
\REF\four{ R. Dijkgraaf, E. Verlinde and H. Verlinde,
On Moduli Spaces of
Conformal Field
Theories with $c \geq 1$, Proceedings Copenhagen Conference,
Perspectives
in String Theory,
edited by P. Di Vecchia and J. L. Petersen,
World Scientific, Singapore, 1988.}
\REF\five{ A. Shapere and F. Wilczek, Nucl.Phys. B320 (1989) 669.}
\REF\six{ M. Dine, P. Huet and N. Seiberg,
Nucl. Phys. B 322 (1989) 301.}
\REF\seven{ J. Lauer, J. Mas and H. P. Nilles, Nucl. Phys. B351 (1991) 353.}
\REF\eight{ K. Kikkawa and M. Yamasaki, Phys. Lett. B149, (1984) 357;
N. Sakai and I. Senda, Prog. Theor. Phys. 75 (1984)692}
\REF\nine{A. Giveon, E. Rabinovici and G. Veneziano,
Nucl. Phys. B322 (1989) 167.}
\REF\ten {W. Lerche, D. L\"ust and N. P. Warner,
Phys. Lett. B231 (1989) 417.}
\REF\eleven {V. S. Kaplunovsky, Nucl. Phys. B307 (1988) 145}
\REF\twelve {L. J.
Dixon, V. S. Kaplunovsky and J. Louis,  Nucl. Phys. B355 (1991) 649.}
\REF\thirteen{L. E. Ibanez, D. L\"{u}st and G. G. Ross, Phys. Lett. B272
(1991)
25.}
\REF\fourteen{L. E. Ibanez and D. L\"{u}st , Nucl. Phys. B382 (1992) 305.}
\REF\fifteen{J. P. Derendinger, S. Ferrara, C. Kounas and
F. Zwirner, Nucl. Phys. B372 (1992) 145, Phys. Lett. B271 (1991)
307.}
\REF\sixteen{ D. Bailin and A. Love, Phys. Lett. B278 (1992) 125;
Phys. Lett. B292 (1992) 315.}
\REF\seventeen{ P. Mayr and S. Stieberger, MPI-Ph/93-07, TUM-TH-152/93
preprint, to appear in Nucl. Phys.}
\REF\eighteen {D. Bailin, A. Love, W. A. Sabra and S. Thomas, SUSX--TH--93/13,
QMW--TH--93/21, Phys. Lett. B to be published}
\REF\nineteen {D. Bailin, A. Love, W. A. Sabra and S. Thomas, SUSX--TH--93/14,
QMW--TH--93/22, Mod. Phys. Lett. A to be published.}
\REF\twenty{M. Spalinski,  Phys.  Lett. B275 (1992) 47;
J. Erler, D. Jungnickel and H. P. Nilles, Phys. Lett. B276 (1992) 303.}
\REF\twentyone{J. Erler and M. Spalinski, preprint, MPI-PH-92-61,
TUM-TH-147-92}
\REF\twentytwo{ T. Kobayashi and  N. Ohtsubo, preprint DPKU--9103.}
\REF\twentythree{M. Spalinski, Nucl. Phys.
B377 (1992) 339.}
\REF\twentyfour{ L. E. Ibanez, J. Mas, H. P. Nilles and F. Quevedo, Nucl.
Phys.
B301 (1988) 157.}
Orbifold compactifications of string theory [\one, \two] possess various
moduli, 
which are background fields corresponding to marginal 
deformations of the underlying conformal field theory including radii 
and angles of the underlying six dimensional torus. The spectrum of the states
for an orbifold theory is invariant under certain discrete transformations on
the
moduli, together with the winding numbers and momenta, referred to as modular
symmetries 
[\three-\ten]. These modular symmetries also manifest themselves in 
the string loop threshold 
corrections [\eleven-\sixteen] that are of crucial 
importance for unification of gauge coupling constants, 
though in this case it is only the transformations on the moduli associated
with the fixed planes of twisted sectors of the theory that are relevant. 
It is these particular symmetries we shall focus on here.

In the case where there are no Wilson lines and in addition all twisted
sector
fixed planes 
are such that the 6-torus $T_6$ can be decomposed into a direct sum
$T_2\bigoplus T_4$ 
with the fixed plane lying in $T_2$, the group
of modular symmetries is a product of $SL(2,Z)$ factors [\twelve] one for each
of the 
$T$ or $U$ moduli associated with the fixed planes. However, when there are 
twisted sectors with 
fixed planes that cannot be decomposed in this way the group of modular
symmetries 
(relevant to the string loop threshold corrections) is in general a product of
congruence subgroups [\seventeen-\nineteen] of
$SL(2, Z).$ It is also known [\twenty, \twentyone, \fourteen] that Wilson
lines
can break $SL(2, Z)$ 
modular symmetries. The present paper is directed towards finding modular
symmetries relevant to string
loop threshold corrections in orbifold theories with Wilson lines, but with 
the simplifying feature that all fixed planes of the twisted sectors allow the
decomposition
of the six torus discussed above.

In the presence of metric, antisymmetric tensor and Wilson line background
fields, 
the momenta $P_L$ and $P_R$ for the left and right movers in the lattice
basis$^*$

\footnote* {for string slope parameter $\alpha'$ taken to be $1\over2$} take
the form 
(see, for example, ref. [20])
$$P_L=\Big({m\over 2} +(g-h)n-{1\over2}A^tCl,\quad  l+An\Big)\equiv 
\Big({\tilde P}_L, \quad l+An\Big)\eqn\school$$
and
$$P_R=\Big({m\over2}- (g+
h)n-{1\over2}A^tCl, \quad 0\Big)\equiv 
\Big({\tilde P}_R,\quad  0\Big)\eqn\east$$
where $m$ and $n$ are the momentum and winding number for the compact
manifold,
$l$ is the momentum on the $E_8\times E'_8$ lattice, $g$, $B$ and $A$ are the
metric, antisymmetric tensor and Wilson line background fields, $C$ is the 
Cartan metric for the $E_8\times E'_8$ lattice, and
$$h=B+{1\over4}A^tCA\eqn\west$$

(Constant shifts in the $E_8\times E'_8 $ lattice due to
point group embeddings of Wilson lines are not relevant for these
purposes [\twenty] and have been suppressed throughout.)
If $\theta$ generates the point group of the orbifold, it is the $\theta^k$
twisted sectors with
fixed planes that are relevant to string loop threshold corrections as
mentioned
earlier. The action of the point group on the winding numbers and momenta can
always
be written in the form [\twenty, \twentyone]
$$u\rightarrow u'=Ru\eqn\square$$ where 
$$u = \pmatrix{n \cr m\cr l}\eqn\egham$$
and
$${\cal R}=\pmatrix{Q&0&0\cr\alpha&Q^*&(1-Q^*)A^tC\cr A(I-Q)&0&I}\eqn\london$$
with
$$\alpha={1\over2}A^tCA(I-Q)+{1\over2}(I-Q^*)A^tCA+2Q^*\delta\eqn\noway$$
where $\delta$ is an antisymmetric integer matrix, which we shall take
to be zero in what follows and * denotes the inverse transpose.
In \london, the matrix $Q$ defines the action of the point group element
$\theta$ on
the basis vectors $e_a$ of the lattice of the six-torus 
$$\theta: \quad e^i_a\rightarrow e^i_b Q_{ba}\eqn\garden$$
The boundary conditions for the $\theta^k$ twisted sector require
$$ P_L=\theta^k P_L, \qquad 
P_R=\theta^k P_R\eqn\paul$$ 
which in terms of the matrix $R$
are 
the conditions
$$R^k u=u.\eqn\paris$$ 
For convenience let the fixed plane of $\theta^k$ be the first complex plane.
Then, $Q^k$ is block diagonal with the $2\times2$ identity matrix as its
leading block. The solution of \paris\ is
$$(I-Q^k)n=0\eqn\horror$$
and
$$(I-{Q^*}^k){\hat p}=0\eqn\pinhead$$
where $$\hat p=m-{1\over2}A^tCAn-A^tCl.\eqn\elm$$
Consequently, $n$ and $\hat p$ can only take non-zero entries for their first
two components. If we use the variables $n$, $\hat p$ and $l+An$, then
the problem is two dimensional so far as $n$ and $\hat p$ are concerned. It
is thus convenient to define
$$u_{\perp}=\pmatrix{n\cr \hat p}\eqn\munich$$
In this basis, the action of the point group element $\theta$ is simply
$$u_{\perp}\rightarrow u'_{\perp}=R_{\perp}u_{\perp},\eqn\george$$
with $l+An$ left invariant by the point group, where
$$R_{\perp}=\pmatrix{Q&0\cr 0&Q^*}\eqn\blackforest$$
In \munich\ and \blackforest, $n$ and $\hat p$ are now understood to be
restricted to their first 2 components and $Q$ to the $2\times 2$ block that
acts on the first complex plane (corresponding to the fixed plane in the
$\theta^k$ twisted sector.)

In terms of the variables $n$, $\hat p$ and $l+An$, the momenta $P_L$ and
$P_R$ take the form
$$P_L\equiv\Big({\tilde P}_L,\quad  l+An\Big)=\Big({{\hat p}\over 2} +(g-B)n,
\quad l+An\Big)\eqn\anne$$
and
$$P_R\equiv\Big({\tilde P}_R,\quad 0\Big)=\Big({\hat p\over 2} -(g+B)n,\quad
0\Big)
\eqn\pat$$
where $g$ and $B$ are now understood to be restricted to the $2\times 2$
block that acts on the first complex plane. The components ${\tilde P}_L$ and
${\tilde P}_R$  on the compact manifold are just as for the case of a 2
dimensional orbifold without Wilson lines except for the replacement of $m$ by
$\hat p.$
The world sheet momentum $P$ is given by
$$\eqalign{P=&{1\over2}{\tilde P}^t_L g^{-1} {\tilde P}_L+{1\over2}
(l+An)^tC(l+An)
- {1\over2}{\tilde P}^t_R g^{-1}
{\tilde P}_R\cr =&{1\over2} u^t_{\perp}\eta u_{\perp}+{1\over2}
(l+An)^tC(l+An)}\eqn\james$$
where
$$ \eta = \pmatrix{0 & {\bf 1}_2 \cr
     {\bf 1}_2 & 0 }\eqn\street$$
and the Hamiltonian $H$ is given by
$$\eqalign{H=&{1\over2}{\tilde P}^t_L g^{-1} {\tilde P}_L+{1\over2}
(l+An)^tC(l+An)
+ {1\over2}{\tilde P}^t_R g^{-1}
{\tilde P}_R\cr =&{1\over2} u^t_{\perp}\Xi u_{\perp}+{1\over2}
(l+An)^tC(l+An)}\eqn\studio$$
where

$$\Xi=\pmatrix{
      2(g-B)g^{-1}(g+B) & Bg^{-1}\cr
	       - g^{-1} B& {1\over2}g^{-1}}\eqn\simon$$

We look for modular symmetries that act only on the components $\tilde P_L$
and $\tilde P_R$ associated with the compact manifold, leave the momentum
$l+An$ associated with the internal $E_8\times E'_8$ degrees of freedom
invariant and leave the Wilson line $A$ invariant.

Consider the transformation
$$u_{\perp}\rightarrow \Omega_{\perp}^{-1}u_{\perp},\eqn\factories$$
where $\Omega_{\perp}$ is a matrix with integer entries. In order to
obtain a well-defined transformation on the integral components of $m$,
$n$ and $l$, it is necessary to place restrictions on the integers occuring in
$\Omega_{\perp}$. We return to this point shortly.
For $\Omega_{\perp}$ to be consistent with level matching it must leave the
world
sheet momentum $P$ invariant and so must satisfy
$$\Omega^t_{\perp}\eta_{\perp}\Omega_{\perp}=\eta_{\perp}.\eqn\mexico$$
The solutions $\Omega_{\perp}$ of \mexico\ form an $O(2,2, Z)$ group
isomorphic to $SL(2, Z)\times SL(2, Z)\times Z_2\times Z_2$
 generated by $\Omega_T$, $\Omega_U$, $\Omega_1$ and $\Omega_2$ where 
$$\Omega_T=\pmatrix{aI&cJ\cr -bJ&dI}\eqn\israel$$
with $$J=\pmatrix{0&1\cr -1&0}\eqn\jordan$$
and $a$, $b$, $c$ and $d$ are integers satisfying
$$ad-bc=1\eqn\europe$$
and
$$\Omega_U=\pmatrix{F&0\cr 0&F^*}\eqn\germany$$
with $$F=\pmatrix{d'&b'\cr c'&a'}\eqn\house$$
and $a'$, $b'$, $c'$ and $d'$ are integers satisfying
$$a'd'-b'c'=1\eqn\live$$
also
$$\Omega_1=\pmatrix{0&0&-1&0\cr0&1&0&0\cr-1&0&0&0\cr0&0&0&1}\eqn\qpr$$
and
$$\Omega_2=\pmatrix{ -1&0&0&0\cr0&1&0&0\cr0&0&-1&0\cr
0&0&0&1}\eqn\swindon$$
The modular symmetries must be compatible with the point group so that
$$\Omega _{\perp}R_{\perp}=R_{\perp}\Omega_{\perp}\eqn\tyre$$
For $Z_N$ orbifolds generated by point group element $\theta$, fixed planes
only occur in $\theta^k$ twisted sectors with $k=2, 3$ or $4$. In other
words, the action of the point group in that complex plane is $Z_2$, $Z_3$
or $Z_4$. Requiring also that the six- torus $T_6$ can be decomposed as 
direct sum $T_2\bigoplus T_4$ with the fixed plane lying in $T_2$, fixed
planes with $Z_2$ point group occur for the $Z_6-II-d$, $Z_8-II-b$ and
$Z_{12}-II$ orbifolds, in the notation of ref. [\twentytwo], fixed planes with
$Z_3$
point group occur for the $Z_3$, $Z_6-I$,  $Z_6-II-b-c-d$ and $Z_{12}-II-b$
 orbifolds, and fixed planes with $Z_4$ point group for the $Z_8-I$
orbifold. For the cases of $Z_2$, $Z_3$ and $Z_4$ fixed planes $Q$ in
\blackforest\
takes the form
$$Q=\pmatrix{-1&0\cr 0&-1},\quad
Q=\pmatrix{0&-1\cr 1&-1},\hbox{and} \ Q=\pmatrix{1&-1\cr 2&-1}\eqn\royal$$
respectively.
Then, \tyre\ is satisfied by both $\Omega_T$ and $\Omega_U$
for $Z_2$ fixed planes, but by only $\Omega_T$ for $Z_3$ and $Z_4$ fixed
planes.
In the $n$, $\hat p$ basis, the moduli associated with fixed planes are
constrained by
$$Q^tgQ=g\eqn\bizzare$$
and $$Q^tBQ=B\eqn\weird$$
for consistency with point group, as in the case without Wilson lines
[\twentythree]
except that $g$ and $B$ are restricted to the $2\times 2$ blocks acting on
the fixed plane. 
If we define the moduli $T$ and $U$ as in the case without
Wilson lines [12]
$$T=T_1+iT_2=2\Big(B_{12}+i\sqrt{\det g}\Big)\eqn\protected$$
and
$$U=U_1+iU_2={1\over g_{11}}\Big(g_{12}+i\sqrt{\det g}\Big).\eqn\shell$$
Then for $Z_2$ fixed planes, both $T$ and $U$ are consistent with the point
group but for $Z_3$ and $Z_4$ fixed planes only $T$ survives as a continuous
modulus and $U$ takes the fixed values
$$U=-{1\over2}(1+i\sqrt{3}),\qquad \hbox{and}\ U=-{1\over2}(1+i)\eqn\mable$$
for $Z_3$
and $Z_4$ respectively.
The $\Xi_{\perp}$ for the Hamiltonian in \studio\ and \simon may be written in
terms
of
the moduli as
$$\Xi_{\perp}={1\over T_2U_2}\pmatrix{W&X\cr Y&Z}\eqn\lebanon$$
with
$$W=\mid T\mid^2\pmatrix{1&U_1\cr U_1&\mid U\mid^2}\eqn\valley$$
$$X=Y^t=T_1\pmatrix{-U_1&1\cr -\mid U\mid^2&U_1}\eqn\generation$$
$$Z=\pmatrix{\mid U\mid^2&-U_1\cr -U_1&1}\eqn\miles$$
Modular transformations $\Omega_{\perp}$ that leave the Hamiltonian invariant
act on the matrix $\Xi_{\perp} $ as
$$\Xi_{\perp}\rightarrow
\Omega_{\perp}^t\Xi_{\perp}\Omega_{\perp}.\eqn\spread$$
It can then be seen that the modular transformations $\Omega_T$ induce the
transformation on the modulus $T$,
$$T\rightarrow {aT+b\over cT+d},\eqn\cult$$
and, in the case of $Z_2$ fixed planes, the modular transformation on the
modulus $U$,
$$U\rightarrow {a'U+b'\over c'U+d'}\eqn\robert$$

As mentioned earlier, it is necessary to put restrictions on the integers
occurring in $\Omega_{\perp}$ in order to insure that \factories\ together
with
$$l+An\rightarrow l+An\eqn\galaxy$$
provide a well defined transformation on the integral components of $m$, $n$
and $l$. In detail, for $\Omega_U$, these conditions are
$$A(I-F^{-1})\in Z\eqn\oxford$$
and $$A^tCA-{1\over2}A^tCAF^{-1}-{1\over2}F^tA^tCA \in Z\eqn\devil$$
Fr $\Omega_T$, the corresponding conditions are
$$cA\in Z \eqn\toff$$
$${c\over2} A^tCA \in Z\eqn\romans$$
$${c} AJA^tC \in Z\eqn\oral$$
$$(1-d)A-{c\over 2} AJA^tCA \in Z\eqn\stone$$
$$(1-a)CA+{c\over 2} CAJA^tCA \in Z\eqn\davros$$  
and
$$(1-{a\over2}-{d\over2})A^tCA-{c\over 4} A^tCAJA^tCA \in Z\eqn\wales$$
The integers $a'$, $b'$, $c'$ and $d'$ in \house\ are then constrained by
\oxford\ 
and \devil, and the integers $a$, $b$, $c$ and $d$ in \israel\ are constrained
by
$\toff $-$\wales $, in a way that depends on the choice of Wilson line $A$.

Acceptable Wilson lines must satisfy [\twenty, \twentyone, \twentyfour]
$$A(I-Q) \in Z$$
and
$${1\over2}A^tCA(I-Q)+{1\over2}(I-Q^*)A^tCA \in Z\eqn\camelot$$
as can be seen from \london\ and \noway, with $\delta=0$
For the case of $Z_2$ fixed plane, a simple example is provided by
$$A^t={1\over2}\pmatrix{0&0&0&0&0&0&1&1\cr 0&0&0&0&0&0&1&1}$$
where we have for simplicity assumed that the embedding is entirely in the
first $E_8$. Only the first 2 rows of $A^t$ have been displayed since the
other 4 rows can not contribute to the non-zero components of $\hat p$.
For this choice of Wilson lines, eqns. $\toff $ -$\wales$ 
imply that the modular
symmetries $\Omega_{T} $ on the $T$ modulus are restricted to the subgroup of
$SL(2, Z)$
characterized by
$$c=0\ (mod 2) ,\qquad a, d= 1\ (mod2).\eqn\arthur$$

On the other hand, \oxford\ and \devil\ require the modular transformations
$\Omega_U$ on the $U$ modulus to be restricted to a subgroup of $SL(2, Z)$
satisfying the constraints
$$a' ,\  d'=0\ (mod2),\qquad b',c'=1\ (mod2)\eqn\saxon$$
or $$a' ,\  d'=1\ (mod2),\qquad b',c'=0\ (mod2).\eqn\saxo$$
For the case of a $Z_3$ fixed plane, a simple example is
$$A^t={1\over3}\pmatrix{1&0&0&0&0&0&1&1\cr 1&0&0&0&0&0&1&1}\eqn\county$$
Then there is only a $T$ modulus and the modular transformations $\Omega_T$
are limited to the subgroup of $SL(2, Z)$ determined by
$$c=0\ (mod 3) ,\qquad a,  \ d= 1\ (mod 3)\eqn\king$$

The main conclusion we can draw about the presence of Wilson lines in 
orbifold models, is that they break some of the duality symmetries 
associated with the moduli of invariant planes. In the  
specific examples considered
above we can see that the $T\rightarrow -1/T $ and 
$U \rightarrow -1/U $ transformations are broken.  A simpler model
with this property is compactification on a  1-dimensional orbicircle. The 
procedure for analysing the duality symmetries of this model including 
Wilson lines, is as given above for the more
realistic case. As before, we look for modular symmetries that act only the  
components of $P_L $ and $P_R $ associated with the orbicircle, and leave
the internal $l + An $ momentum and the Wilson line $A$ invariant.
The world sheet momentum $P$ and  
Hamiltonian $H$ are given by eqns $\james $ and $\studio $ with 
$$ \eta = \pmatrix{0 & { 1} \cr
     {1}& 0 }\eqn\toe$$
An orbicircle of radius $R$ is constructed from a circle of the 
same radius, by modding out by a ($Z_2 $-valued ) reflection symmetry.
The matrix $\Xi $ in this case is given by 
$$\Xi=\pmatrix{
      2{R^2} & 0 \cr
	       0 & {1\over{\displaystyle 2  R^2} } }\eqn\slimy$$
Duality symmetries leaving $P$ invariant take the form  
$$\Omega = \pmatrix{0 & \rho \cr 1\over{\displaystyle \rho} & 0 } \eqn\me $$
where $\rho $ is a real parameter. From the constraints on $\rho $, 
 ( analogous to those listed in eqns $\toff $ - $\wales $ )
one may easily deduce that  
$${1\over{\rho }}  \in Z\eqn\dot $$
and
$${1\over{\displaystyle 2 \rho }} (A^t CA ) \in Z \eqn\bisc$$
The constraint \dot\  fixes $\rho = 1$. On the other hand
 since the 
matrix $Q = -1 $ in this case, $A^t C A $ is  half- 
integer valued in general, so constraint \bisc\ is not
satisfied. In fact the only solution to \dot\ and \bisc\
is when $A = 0 $
  i.e. the case 
without Wilson lines, which corresponds to the usual  
$2R^2 \rightarrow 1/{2 R^2 }$ duality symmetry. Therefore, even in this 
relatively simple example, we can see how Wilson lines break the 
`stringy' duality that exchanges the radius with inverse radius.

In conclusion, we have found a class of target space modular symmetries
relevant to string loop threshold corrections where only
the components ofthe momenta $P_L$ and $P_R$ associated
with the compact manifold transform
while the components in the $E_8\times E'_8$ internal space
and the Wilsonlines are left invariant.
These symmetries are subgroups of $SL(2, Z)$
acting on the moduli $T$ and
$U$ with the specific subgroup
being determined by the form of the Wilson lines.
\vskip 3cm
\centerline{\bf{ACKNOWLEDGEMENT}}
This work was supported in part by S.E.R.C. and the work of S. T. was
supported by the Royal Society.
We would like to thank L. E.  Ibanez
for helpful communications.
\refout
\end